\documentclass{ws-procs9x6}

\newcommand{\be}{\begin{equation}}
\newcommand{\ee}{\end{equation}}
\newcommand{\bea}{\begin{eqnarray}}
\newcommand{\eea}{\end{eqnarray}}
\newcommand{\PR}[1]{{\it Phys.\ Rev.}\ {\bf #1}}
\newcommand{\PRL}[1]{{\it Phys.\ Rev.\ Lett.}\ {\bf #1}}
\newcommand{\EPJ}[1]{{\it Eur.\ Phys.\ J.}\ {\bf #1}}
\newcommand{\PL}[1]{{\it Phys.\ Lett.}\ {\bf #1}}
\newcommand{\NP}[1]{{\it Nucl.\ Phys.}\ {\bf #1}}

\newcommand{\bfk}{\mbox{\boldmath $k$}}

\newcommand{\bfP}{\mbox{\boldmath $P$}}
\newcommand{\bfS}{{\mbox{\boldmath $S$}}_{_T}}

\newcommand{\bfSL}{{\mbox{\boldmath $S$}}_{_L}}
\newcommand{\bfsL}{{\mbox{\boldmath $s$}}_{_L}}

\newcommand{\pup}{p^\uparrow}
\newcommand{\pdown}{p^\downarrow}

\def\bkt{\bf k_\perp}
\def\bpt{\bf p_\perp}
\def\kt{k_\perp}
\def\pt{p_\perp}

\begin{document}

\title{Including Cahn and Sivers effects into event generators}

\author{A.~Kotzinian}

\address{Dipartimento di Fisica Generale, Universit\`a di Torino\\
         and \it INFN, Sezione di Torino, Via P. Giuria 1, I-10125 Torino, Italy\\
and \it Yerevan Physics Institute, 375036 Yerevan, Armenia \\
and \it JINR, 141980 Dubna, Russia\\
E-mail: aram.kotzinian@cern.ch}

\maketitle

\abstracts{It is demonstrated that event generators {\tt LEPTO} and
{\tt PYTHIA} can be modified to describe some azimuthal modulations.
The comparisons of results obtained with modified {\tt LEPTO} with
existing data in the current fragmentation region of SIDIS are
presented for Cahn and Sivers effects as well as the predictions for
the target fragmentation region. The predictions for Sivers effect
in the Drell--Yan process obtained with modified {\tt PYTHIA} are
also presented. The concept of hadronization function is discussed.}

\section{Introduction\label{sec:intro}}

The spin (in)dependent azimuthal asymmetries arising in high energy
reactions allow us to study the dynamical effects related to spin
and transverse momentum of partons in target nucleon and in
hadronization.

The Drell--Yan lepton pair production is the simplest process which
does not include the hadronization dynamics and within QCD can be
described using as a nonperturbative input only the parton
distribution functions\footnote{In the following the notations
of\cite{akp1} are used.}:
\be
    \label{dy_sym}
        d\sigma^{h_1+h_2 \to \ell^+ + \ell^- + X }
        \sim \sum_q  (f_{{\bar q}/h_1}f_{q/h_2}+1\leftrightarrow2)
    \; \otimes d\hat\sigma^{q+\bar q \to \ell^+ + \ell^-}.
\ee
The predictions for the Sivers\cite{siv} effect for this process
recently have been presented in\cite{akp2,vogyuan}. Here it will be
demonstrated that similar results are obtained using modified {\tt
PYTHIA} event generator.

More nonperturbative inputs are needed to describe the
semi-inclusive DIS (SIDIS) processes within the QCD formalism.
Namely, for the particles produced in the current fragmentation
region (CFR) of SIDIS one needs to introduce fragmentation function,
$D_{q}^h(z)$:
\be
    \label{sidis_cfr}
    d\sigma^{\ell p \to \ell h X } \sim \sum_q  f_{q} \otimes
    d \hat\sigma ^{\ell q\to \ell q} \otimes D_q^h.
\ee
For the particles produced in the target fragmentation region (TFR)
another nonperturbative input --- the fracture functions,
$M_{q/N}^h(x,z)$, are needed:
\be
    \label{sidis_tfr}
    d\sigma^{\ell p \to \ell h X } \sim \sum_q
    d \hat\sigma ^{\ell q\to \ell q} \otimes M_{q/N}^h.
\ee
In practice it is not easy to separate the two regions at moderate
beam energies and final hadronic state invariant
masses\cite{berger}. An alternative approach which is able to
describe the particles production in the whole phase space is based
on the LUND string fragmentation model and adopted in the Monte
Carlo event generators\cite{lepto,JETSET}. Here the SIDIS cross
section can be represented as\cite{ak2}
\be
    \label{sidis_hf}
    d\sigma^{\ell p \to \ell h X } \sim \sum_q  f_{q} \otimes
    d \hat\sigma ^{\ell q\to \ell q} \otimes H_{q/N}^h,
\ee
where the hadronization function, $H_{q/N}^h$, describs the particle
production from the system formed by the struck quark and target
remnant.

In\cite{ak1,akp1} the role of parton intrinsic motion in SIDIS
processes in CFR within QCD parton model has been considered at
leading order; intrinsic $\bkt$ is fully taken into account in quark
distribution functions and in the elementary processes as well as
the hadron transverse momentum, $\bpt$, with respect to fragmenting
quark momentum.

The average values of $\kt$ for quarks inside protons and $\pt$ for
final hadrons inside the fragmenting quark jet where fixed by a
comparison with data on Cahn effect\cite{cahn} -- the dependence of
the unpolarized cross section on the azimuthal angle between the
leptonic and the hadronic planes. The single spin asymmetry (SSA)
$A_{UT}^{\sin(\phi_\pi - \phi_S)}$ recently observed by
HERMES\cite{hermUT} and COMPASS\cite{compUT} Collaborations was
successfully described by Sivers mechanism.

Here it will be demonstrated that both Cahn and Sivers effects can
be implemented into Monte Carlo event generators.

\section{Including Cahn effect in LEPTO\label{sec:cahn}}

In the simplest case, corresponding to LO approximation of parton
model, event generation in {\tt LEPTO} proceeds in several steps:
\begin{enumerate}
\item the hard scattering kinematics is generated,
\item the active quark inside the nucleon is chosen
according to the quark density function $f_q(x,Q^2)$,
\item the transverse momentum of the final quark is simulated with Gaussian $\kt$
and flat $\varphi$ distributions. Note that the transverse momentum
of the final final quark is equal to that of initial quark for
leading order hard subprocess.
\item the string fragmentation machinery of {\tt JETSET}
program\cite{JETSET} is applied to form the final hadrons.
\end{enumerate}

The Cahn effect\cite{cahn} is a kinematical effect arising due to
the presence of nonzero intrinsic transverse momentum of quarks in
the nucleon. In the general case of non collinear kinematics
Mandelstam variables depend on the quark transverse momentum and its
azimuthal angle and at order ${\mathcal O}(\kt/Q)$ one has

\be
    d\hat\sigma^{\ell q\to \ell q} \propto
    1-\frac{(2+y)\sqrt{1-y}}{1+(1-y)^2}\frac{\kt}{Q}\cos\varphi.
    \label{lqcahn}
\ee
Eq. (\ref{lqcahn}) shows that the azimuthal angle of the final
quark (and of the string's end associated with the struck quark)
is now modulated with amplitude depending on $y, Q$ and $\kt$.

This effect can be introduced in the {\tt LEPTO} event generator at
the step (3) of the event generation, when the transverse momentum
and azimuthal angle of the scattered quark are generated. To do this
the generation of the quark transverse momentum, $\kt$, is left
unchanged and then the azimuthal angle is generated according to Eq.
(\ref{lqcahn}). This leads to azimuthal modulation of the string
axis. The momentum conservation means that the transverse momentum
of the quark is balanced by that of the target remnant, which in
turn means that the azimuthal angle of the target remnant
$\varphi_{qq}=\varphi+\pi$. Hence, one expects that the azimuthal
angle of the hadrons in the target fragmentation region (TFR),
$x_F<0$, will be modulated with a phase shifted by $\pi$ with
respect to that in CFR.
\begin{figure}[h!]
\begin{center}
\vspace {-0.5cm}
 \includegraphics[width=0.48\linewidth, height=0.45\linewidth]{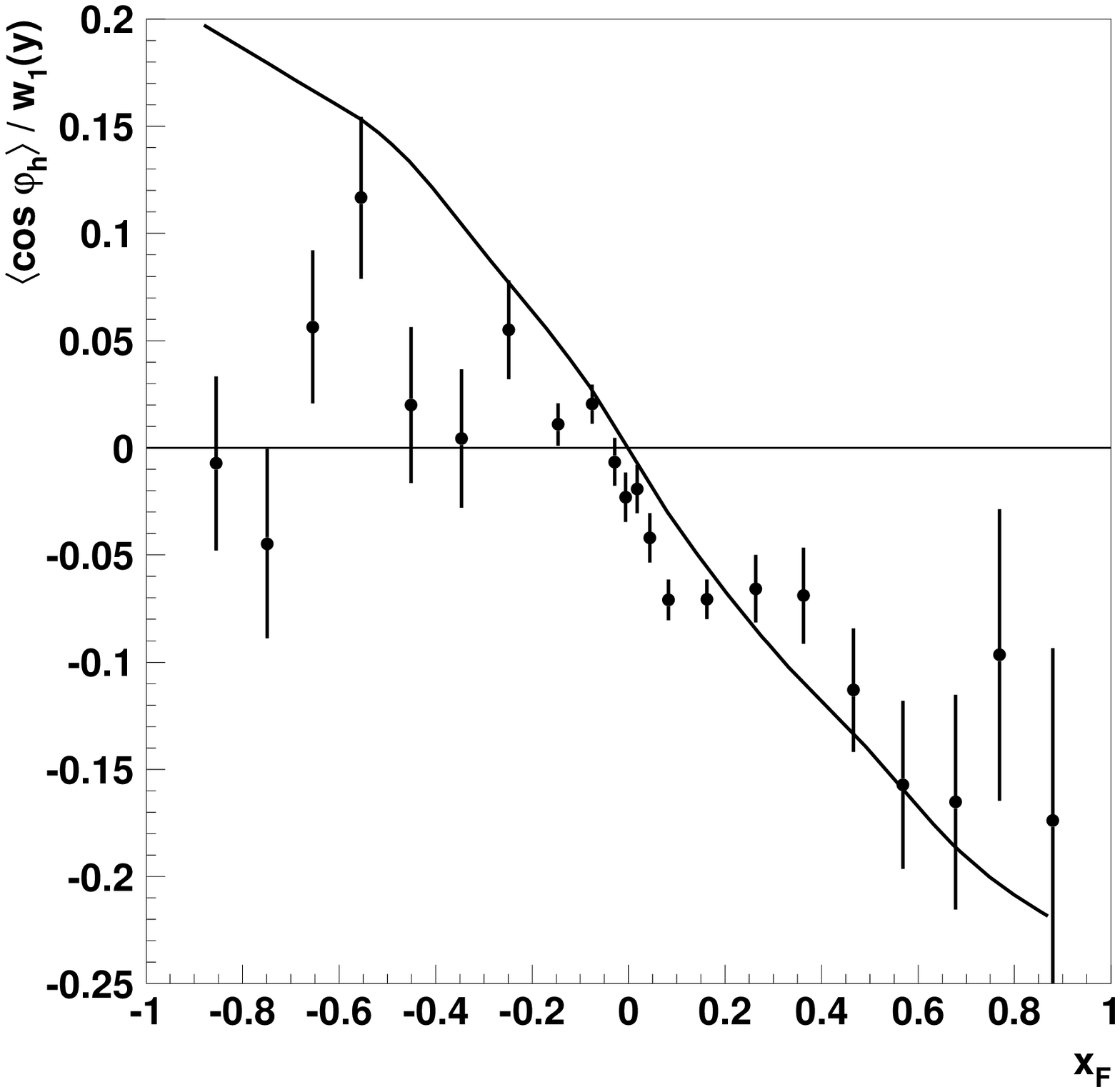}
\hfill
 \includegraphics[width=0.48\linewidth, height=0.45\linewidth]{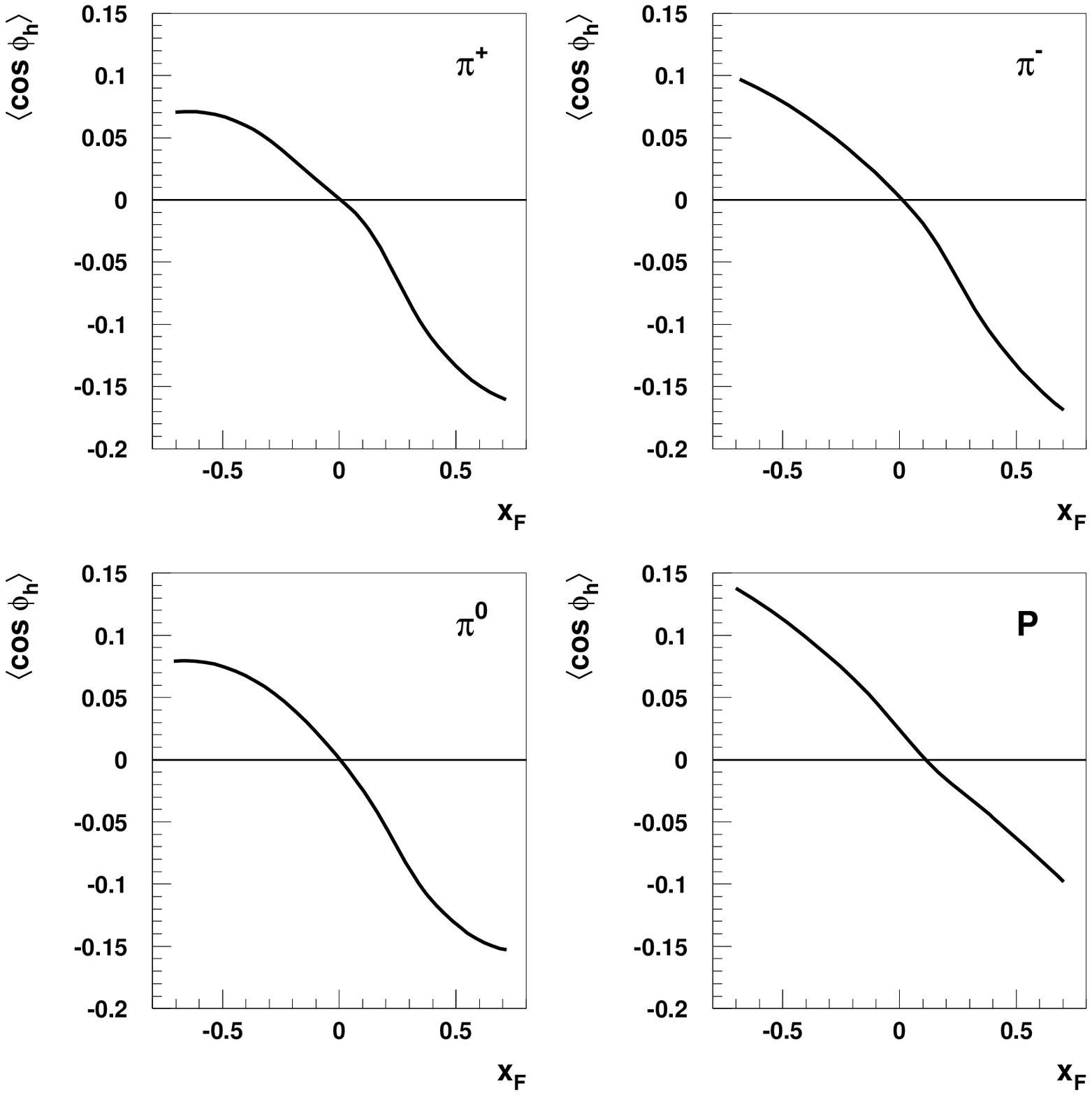}
\caption{\label{fig:cahn} {Left: The $x_F$ dependence of $\langle
\cos \phi_h \rangle / w_1(y)$ for charged hadrons compared with EMC
data. Right: Predictions of modified {\tt LEPTO} for $x_F$
dependence of $\langle \cos \phi_h \rangle$ for different hadrons
produced in 12 GeV unpolarized SIDIS process.}}
\end{center}
\vspace {-0.5cm}
\end{figure}

Data on azimuthal dependencies of SIDIS covering a large $x_F$ range
have been obtained by the EMC Collaboration\cite{emc} for a beam
energy of 280 GeV. The $x_F$ dependence of $\langle \cos \phi_h
\rangle /w_1(y),$ where
$w_1(y)=(2-y)\sqrt{1-y}/\left(1+(1-y)^2\right)$, obtained by using
modified {\tt LEPTO} for EMC kinematics are presented in
Fig.~\ref{fig:cahn} together with data points from\cite{emc} (left
panel). The simulations has been done with LO setting of {\tt LEPTO}
(LST(8)=0) and with values of the parameters describing intrinsic
$k_T$ (PARL(3)=0.5) and fragmentation $p_T$ (PARL(21)=0.45) as
adopted in\cite{akp1}.

The predictions of modified {\tt LEPTO} for $\langle \cos \phi_h
\rangle$ of different hadron ($\pi^+$, $\pi^-$, $\pi^0$ and $p$)
produced in SIDIS on a proton target at future CEBAF 12 GeV facility
at JLab\cite{jlab12} are presented in Fig.~\ref{fig:cahn} (right
panel). One can see from Fig.~\ref{fig:cahn} and that the predicted
mean value of $\cos \phi_h$ in the CFR is negative $\langle \cos
\phi_h \rangle_{CFR}<0$, while in the TFR is positive $\langle \cos
\phi_h \rangle_{TFR}>0$, as suggested by arguments based on
transverse momentum conservation.

\section{Including Sivers effect in LEPTO\label{sec:sivers}}

The azimuthal modulation of the string transverse momentum in the
previous section was due to Cahn effect -- the dependence of the
non planar lepton-quark scattering cross section on the quark
azimuth. The quark distribution, $f_q(x,k_\perp)$ itself is
independent of quark azimuthal angle.

The situation is different when one considers SIDIS on a
transversely polarized nucleon. Now a correlation between transverse
momentum of quark, $\bfk_\perp$ and target transverse polarization
$\bfS$ of type ${\bfS} \cdot [\hat {\bfP} \times \hat{\bfk}_\perp]$
is possible -- the so called Sivers effect\cite{siv}.

The unpolarized quark (and gluon) distributions inside a
transversely polarized proton can be written as:
\be
    f_ {q/\pup} (x,\bfk_\perp) = f_ {q/p} (x,\kt) + \frac{1}{2} \,
    \Delta^N f_ {q/\pup}(x,\kt)  \; {\bfS} \cdot (\hat {\bfP} \times
    \hat{\bfk}_\perp)\; . \label{poldf}
\ee
Eq. (\ref{poldf}) implies
\bea
    \nonumber &&f_ {q/\pup} (x,\bfk_\perp) + f_ {q/\pdown}
    (x,\bfk_\perp) =
    2 f_ {q/p} (x,\kt)\;, \\
    &&f_ {q/\pup} (x,\bfk_\perp) - f_ {q/\pdown} (x,\bfk_\perp) =
    \Delta^N f_ {q/\pup}(x,\kt)\;{\bfS} \cdot (\hat{\bfP}  \times \hat
    {\bfk}_\perp)\;, \label{sivf}
\eea
where $f_ {q/p} (x,\kt)$ is the unpolarized parton density and
$\Delta^N f_ {q/\pup}(x,\kt)$ is referred to as the Sivers function.
Notice that, as requested by parity invariance, the scalar quantity
$\bfS \cdot (\hat{\bfP} \times \hat {\bfk}_\perp)$ singles out the
polarization component perpendicular to the $\bfP-\bfk_\perp$ plane.
For a proton moving along $-z$ and a generic transverse polarization
vector $\bfS = |\bfS|\,(\cos\phi_S, \sin\phi_S, 0)$ one has:
\be
    \bfS \cdot (\hat{\bfP}  \times \hat{\bfk}_\perp) = |\bfS| \,
    \sin(\varphi-\phi_S) \equiv |\bfS| \, \sin\phi_{Siv} \,,
\ee
where $(\varphi-\phi_S) = \phi_{Siv}$ is the Sivers angle.

The Sivers function for each light quark flavor $q=u,d$ are
parameterized in the following factorized form\cite{akp1}:
\be
    \Delta^N f_ {q/\pup}(x,\kt) = 2 \, {\mathcal N}_q(x) \, h(\kt) \,
    f_ {q/p} (x,\kt)\; , \label{sivfac}
\ee
where
\bea
    &&{\mathcal N}_q(x) =  N_q \, x^{a_q}(1-x)^{b_q} \,
    \frac{(a_q+b_q)^{(a_q+b_q)}}{a_q^{a_q} b_q^{b_q}}\; ,
    \label{siversx} \\
    &&h(\kt) = \sqrt{2e} \, \frac{\kt} {M} \, e^{-\kt^2/M^{2}}\;  ,
    \label{siverskt}
\eea
where $N_q$, $a_q$, $b_q$ and $M$ (GeV/$c$) are parameters. Then
Eq.~(\ref{poldf})can be rewritten as
\be
    f_ {q/\pup} (x,\bfk_\perp) = f_ {q/p} (x,\kt)[1+|\bfS|{\mathcal N}_q(x)h(\kt) \,
    \sin\phi_{Siv}].
    \label{sivmod}
\ee

Again, the Sivers effect is incorporated into {\tt LEPTO} at the
stage 3) of the event generation in the same way as for the Cahn
effect but now the azimuthal angle is generated according to
Eq.~(\ref{sivmod}). For simulations the following set of parameters
compatible with those obtained in\cite{akp1,akp2} have been used:
$N_u=N_{\bar u}=0.5$, $N_d=N_{\bar d}=-0.2$, $a_q=0.3$, $b_q=2$ and
$M^2=0.36$ (GeV/c)$^2$.

\begin{figure}[h!]
\begin{center}
\vspace {-0.5cm}
 \includegraphics[width=0.48\linewidth, height=0.45\linewidth]{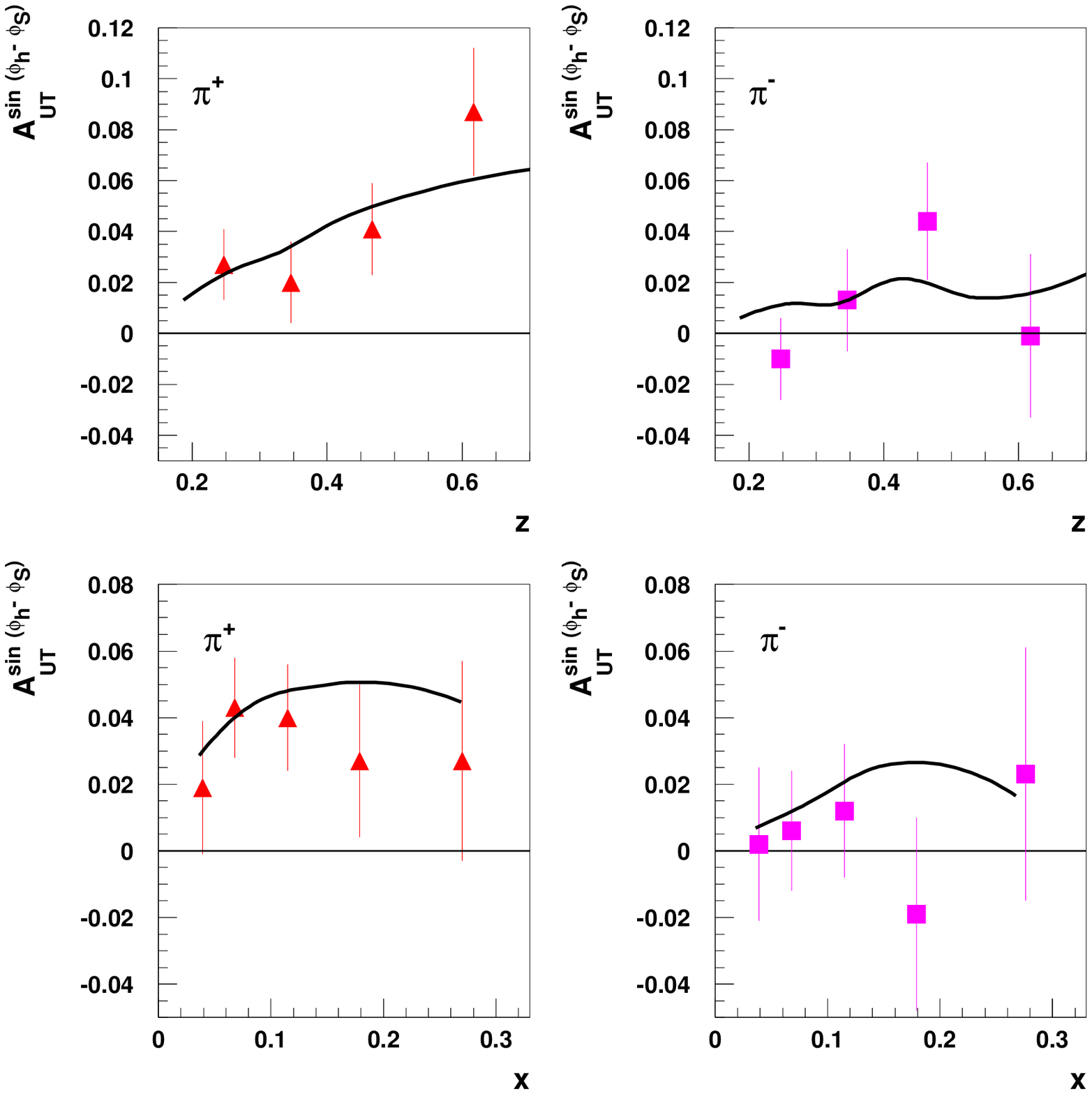}
\hfill
 \includegraphics[width=0.48\linewidth, height=0.45\linewidth]{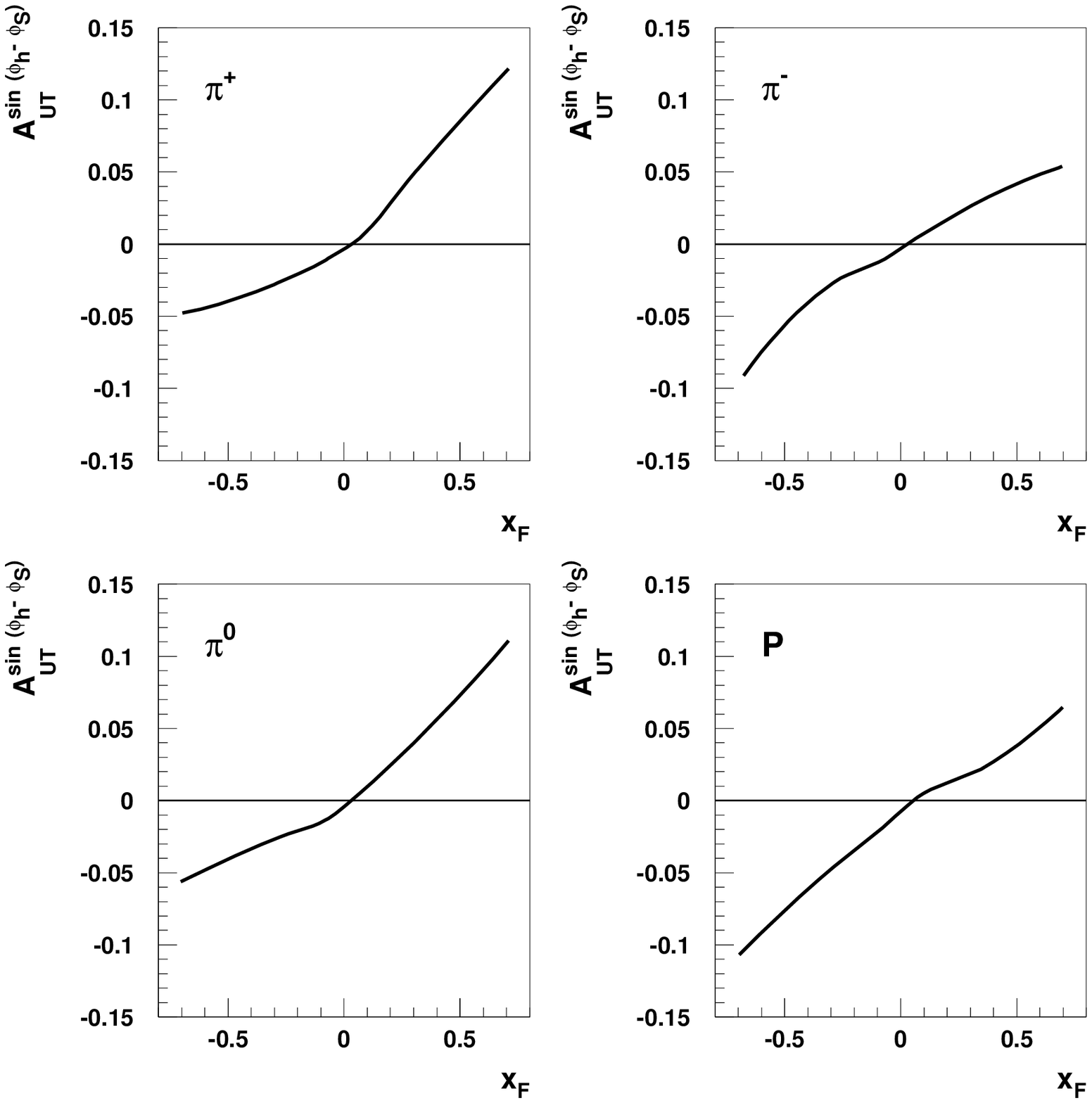}
\caption {Left: HERMES data on $A_{UT}^{\sin(\phi_\pi-\phi_S)}$
for scattering off a transversely polarized proton target. The
curves are the results of simulations obtained with modified {\tt
LEPTO}; Right: Predicted dependence
    of $A_{UT}^{\sin(\varphi_h-\varphi_S)}$ on $x_F$ for different
    hadrons produced in SIDIS of 12 GeV electrons off a transversely polarized
    proton target.} \label{fig:sivsidis}
\end{center}
\vspace {-0.5cm}
\end{figure}

In Fig.~\ref{fig:sivsidis} the results of simulation for HERMES
experimental conditions are compared with observed Sivers
asymmetries\cite{hermUT} (left panel).

\begin{figure}[h!]
\begin{center}
\vspace {-0.5cm}
 \includegraphics[width=0.48\linewidth, height=0.45\linewidth]{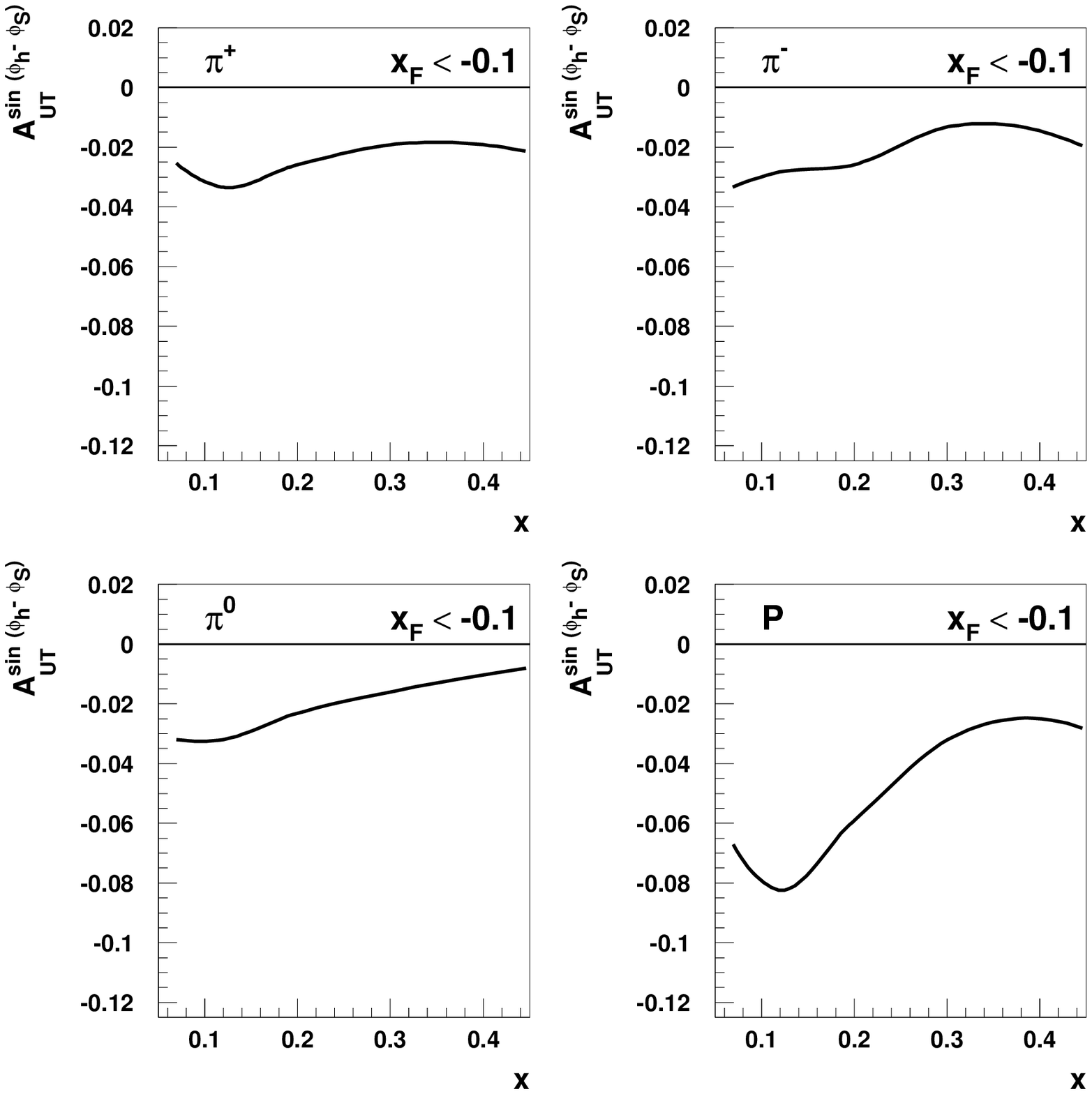}
\hfill
 \includegraphics[width=0.48\linewidth, height=0.45\linewidth]{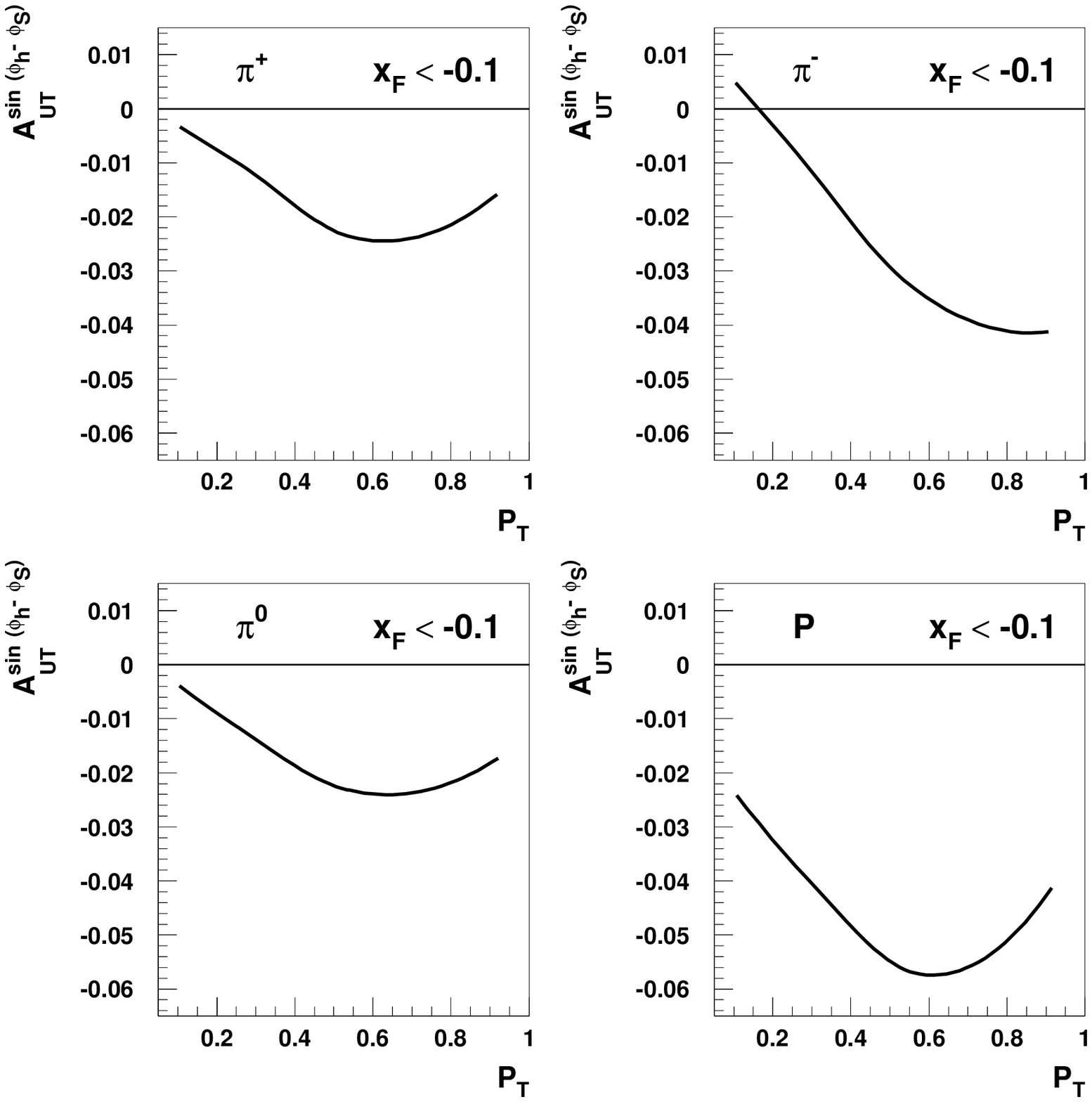}

 \caption{\label{fig:sivxpt} {Predicted dependence
    of $A_{UT}^{\sin(\varphi_h-\varphi_S)}$ on $x$, left panel, ($p_T$, right panel) for different
    hadrons produced in the TFR ($x_F<-0.1$) of SIDIS of 12 GeV electrons off
    a transversely polarized proton target.}}
\end{center}
\vspace {-0.5cm}
\end{figure}

Future facilities as Electron Ion Colliders or upgraded JLab will
have larger kinematic coverage and will offer the possibility of
studying the Sivers effect also with hadrons produced in the TFR. As
an example, the simulations have been done for 12 GeV electron SIDIS
of a proton target. The DIS cut $Q^2>1$ (GeV/c)$^2$ and $W^2>4$
GeV$^2$ and a cut on the produced hadron transverse momentum
$P_T>0.05$ GeV/c was imposed. The predictions for $x_F$ dependence
for JLab kinematics is presented in Fig.~\ref{fig:sivsidis} (right
panel). The  $x$ and $P_T$ dependencies in the TFR are presented
Fig.~\ref{fig:sivxpt}.

\begin{figure}[h!]
\begin{center}
\vspace {-0.5cm}
 \includegraphics[width=0.48\linewidth, height=0.4\linewidth]{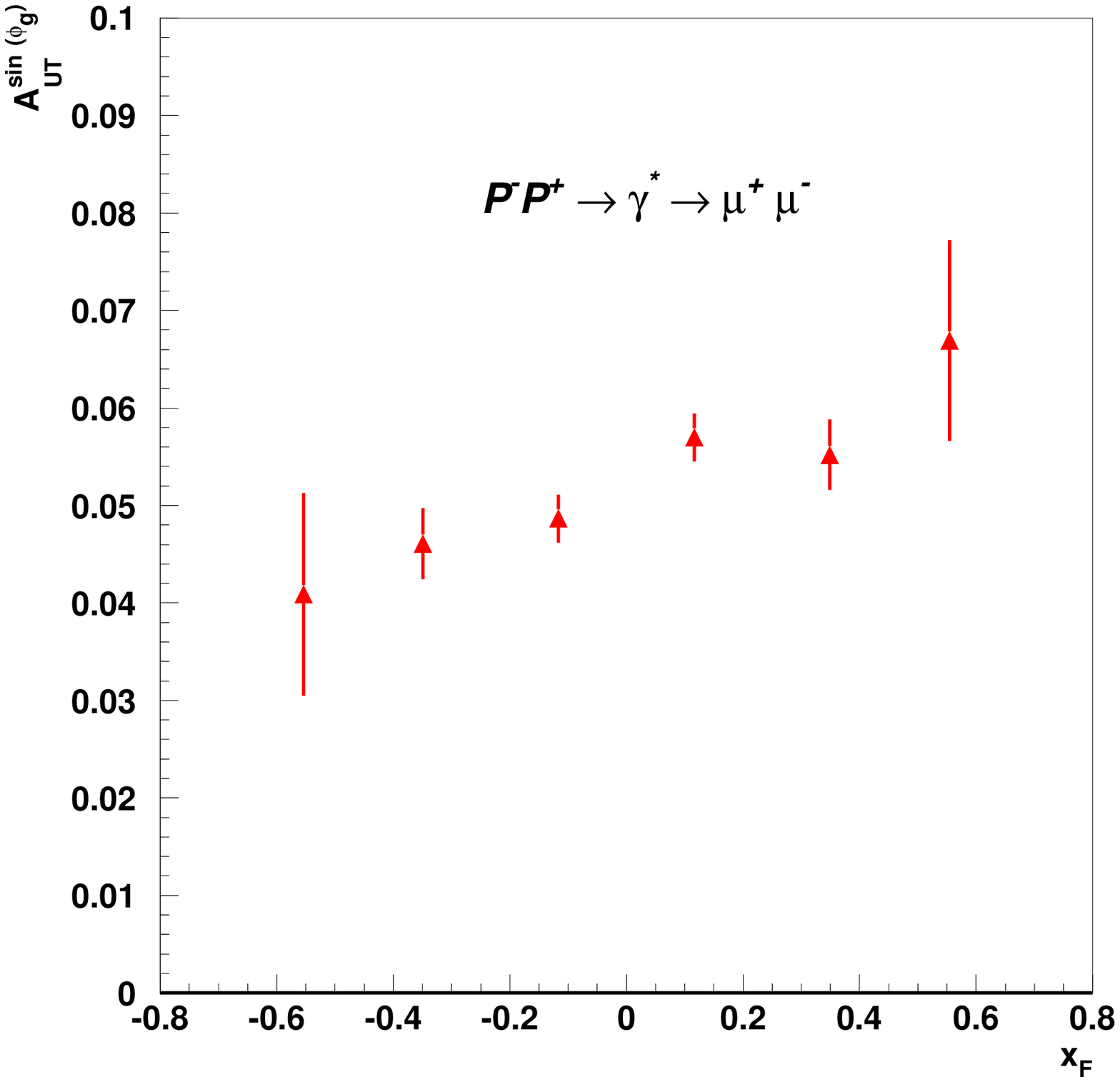}
\hfill
 \includegraphics[width=0.48\linewidth, height=0.4\linewidth]{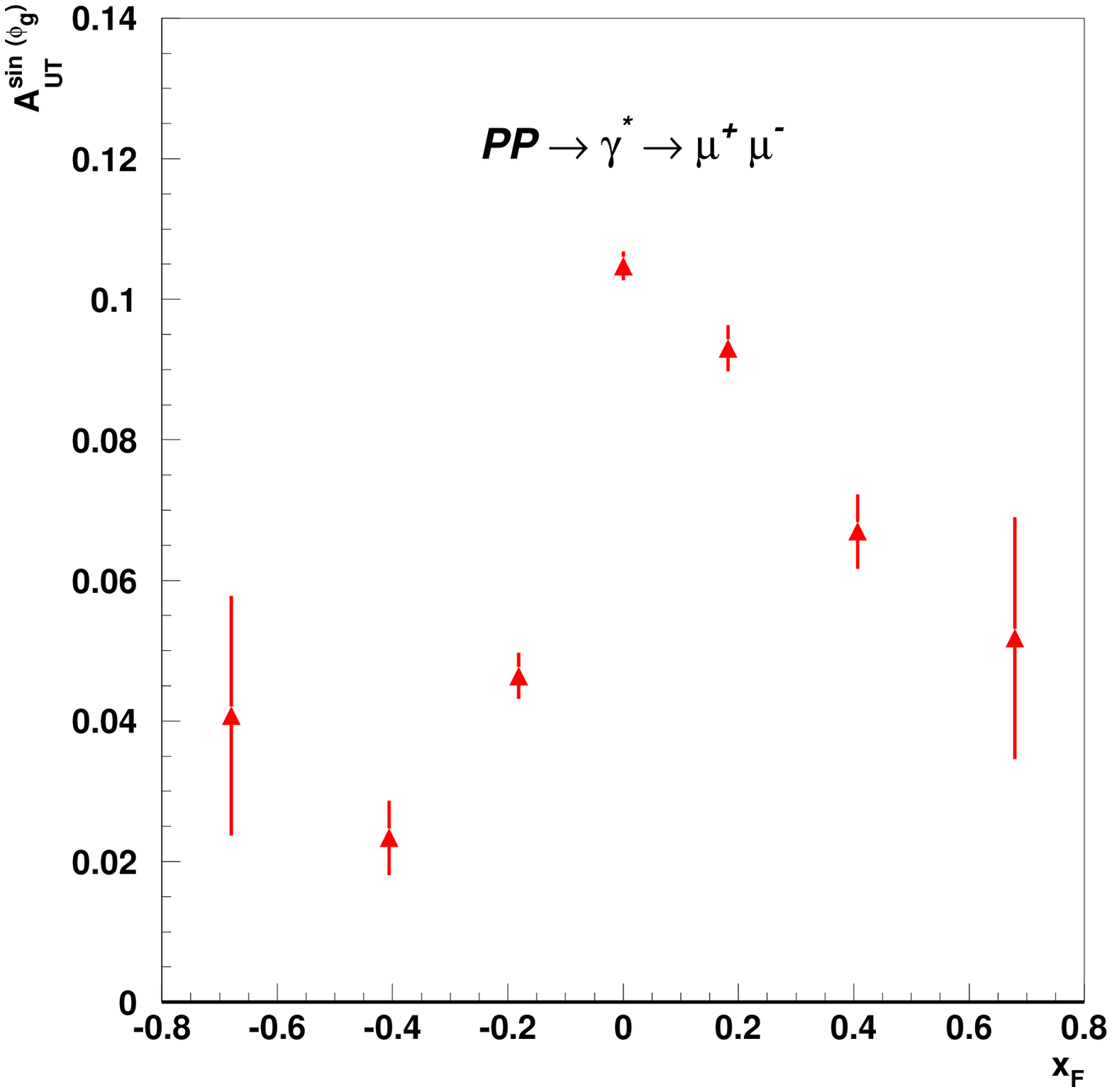}

 \caption{Predicted dependence of $A_N$ for Drell--Yan process on $x_F$.
 Left: $\sqrt{s}=14.4$ GeV, right: $\sqrt{s}=200$ GeV}
    \label{fig:dy}
\end{center}
\vspace {-0.5 cm}
\end{figure}

In Fig.~\ref{fig:dy} the results for Sivers asymmetry in the Drell
--YAn process obtained with modified {\tt PYTHIA} generator are
presented for two different energies: planned GSI $\bar p p$
collider with $\sqrt{s}=14.4 GeV$ (left panel) and RHIC
$\sqrt{s}=200 GeV$ (right panel). The results\footnote{Note, that
here the sign convention of\cite{vogyuan} is adopted.} are similar
to that obtained in\cite{akp2,vogyuan}.

\section{Discussion and Conclusions\label{sec:concl}}

The advantage of this MC based approach compared to standard QCD
factorized approach is the full coverage of produced hadron phase
space. Figs.~\ref{fig:cahn} and~\ref{fig:sivsidis} demonstrate that
the modified {\tt LEPTO} event generator well describs the data in
the CFR both for Cahn and Sivers asymmetries. One can notice in
Fig.~\ref{fig:cahn} that the integrated experimental value of
$\langle \cos \phi_h \rangle$ for charged hadrons in the CFR is not
compensated by that in TFR. It seems improbable that this imbalance
can be compensated by larger values of $\langle \cos \phi_h \rangle$
of neutral hadrons at $x_F \simeq -1$.

Note, that in present approach a possible modifications of
hadronization in the case of polarized target have been ignored.
In\cite{ak2} it was shown that the hadronization functions in
principle depends on polarization states of struck quark and target
nucleon even for production of (pseudo)scalar or unpolarized
particles. This dependence cannot be neglected at moderate energies
and new nonperturbative input
--- the polarized hadronization functions, $\Delta H_{q/N}^h$, are needed to describe the
polarized SIDIS. The expression for the SIDIS helicity asymmetry is
then looks as:

\begin{equation}
  \label{eq:a1fracf}
  A_1^{h}(x,z,Q^2) =
  \frac{\sum_q e_q^2 \,q(x,Q^2)H_{q/N}^h(x,z,Q^2)
  [\frac{\Delta q(x,Q^2)}{q(x,Q^2)}+
  \frac{\Delta H_{q/N}^h(x,z,Q^2)}{H_{q/N}^h(x,z,Q^2)}]}
  {\sum_{q} e_{q}^2 \,q(x,Q^2)\,H_{q/N}^h(x,z,Q^2)
  [1+\frac{\Delta q(x,Q^2)\,
  \Delta H_{q/N}^h(x,z,Q^2)}{q(x,Q^2)\,H_{q/N}^h(x,z,Q^2)}]}.
\end{equation}

Since the hadronization functions depend on target nucleon, active
quark and produced hadron variables the new correlations as ({\bf
a}): ${\bfSL} \cdot [{\bpt} \times \hat{\bfk}_\perp]$, ({\bf b}):
${\bfsL} \cdot [{\bpt} \times \hat{\bfk}_\perp]$, ({\bf c}):
$[{\bfS} \times {\bpt}] \cdot [{\bf p}_T^h \times \hat{\bf s}_T]$
{\it etc} are possible. This correlations cannot be present
separately in the distribution and fragmentation functions. They
will induce the azimuthal asymmetries in the SIDIS of ({\bf a}):
unpolarized lepton off longitudinally polarized target, ({\bf b}):
longitudinally polarized lepton off unpolarized target, ({\bf c}):
unpolarized lepton off transversely polarized target {\it etc}.

The new high statistic measurements in both CFR and TFR of SIDIS
will allow to check the predictions of the approach presented here
and better understand the effects of the quark intrinsic transverse
momentum and hadronization mechanism in SIDIS. The study of single
spin asymmetries in Drell-Yan process will provide an additional
test of our understanding od spin dependent phenomena.

\section*{Acknowledgements}

The author express his gratitude to M.~Anselmino, A.~Prokudin for
discussions and to P. Ratcliffe for the kind invitation to
Transversity 2005 workshop.


\end{document}